\documentclass{JHEP3}
\usepackage{epsfig}

\newcommand{\be}{\begin{equation}}
\newcommand{\ee}{\end{equation}}
\newcommand{\bea}{\begin{eqnarray}}
\newcommand{\eea}{\end{eqnarray}}
\newcommand{\Tr}{{\rm Tr}}
\newcommand{\tr}{{\rm tr}\ }

\title{The large $N$ limit of
four dimensional Yang-Mills field coupled to adjoint fermions on
a single site lattice}
\author{A. Hietanen and R. Narayanan 
\\Department of Physics, Florida International University, Miami,
FL 33199, USA\\E-mail: \email{rajamani.narayanan@fiu.edu}}

\abstract {We consider the large $N$ limit of 
four dimensional $SU(N)$ Yang-Mills field coupled
to adjoint fermions on a single site lattice. We use perturbative techniques
to show that the $Z^4_N$ center-symmetries are broken with na\"ive
fermions but they are not broken with overlap fermions. We use numerical
techniques to support this result. Furthermore, we present evidence
for a non-zero
chiral condensate for one and two Majorana flavors at one
value of the lattice gauge coupling.
}

\keywords{1/N Expansion, Eguchi-Kawai reduction, Adjoint fermions, Lattice Gauge Field Theories}

\preprint{}

\begin{document}

\section{Introduction}
Continuum reduction~\cite{Narayanan:2003fc} holds in the 't Hooft limit of
large $N$ gauge theories in $d>2$ enabling one to
obtain physical results in the infinite volume limit
and zero temperature 
by working on a $l^d$ torus with $l$ of the order
of one or two fermi.  Reducing $l$ below
a certain critical value will force the theory to go into
the deconfined phase.  This in contrast to the theory in $d=2$
where one can set $l=0$ and work on a single site lattice.
This is referred to as Eguchi-Kawai reduction~\cite{Eguchi:1982nm}
and it works in $d=2$ since the physical theory is in the
confined phase for all temperatures.

A weak coupling analysis~\cite{Bhanot:1982sh} on a single site
lattice shows that the large $N$ limit of $SU(N)$ Yang-Mills theory
breaks all $Z^d_N$ symmetries associated with the Polyakov loops
if $d>2$.  Fermions in the fundamental representation
will not affect this argument but fermions in the adjoint
representation can modify the results of the weak coupling
analysis. A continuum analysis of the theory with adjoint fermions
on ${\bf R}^3\times S^1$
with periodic boundary conditions for fermions
in the compact direction shows that the $Z_N$ symmetry
is not broken in that direction~\cite{Kovtun:2007py}. 
An analysis on $S^3\times S^1$ also shows a region where
the $Z_N$ symmetry is not broken~\cite{Hollowood:2009sy} .
A lattice
analysis of the same theory with Wilson fermions indicates
that one can reduce the compact direction to a single site
on the lattice and still maintain the $Z_N$ symmetry
\cite{Bringoltz:2009mi,Bringoltz:2009fj,Poppitz:2009fm}.~\footnote{There is a related
paper on this subject~\cite{Bedaque:2009md} and the reader is
referred to~\cite{Bringoltz:2009fj,Poppitz:2009fm} for a clarification of
possible discrepancies.}  

The natural question that follows is if the $Z_N^d$ symmetries
remain unbroken on a single site lattice in $d>2$ dimensions.
Recent numerical analysis with Wilson fermions in $d=4$ has shown this
is most likely the case for a wide range of quark masses~\cite{Bringoltz:2009kb}.

In this paper we address the same question as above
in $d=4$ using
na\"ive fermions and overlap fermions. The reason to study
na\"ive fermions is the apparent absence of doublers on
a single site lattice since we only have the zero momentum
mode with periodic boundary conditions. But, we will show
using weak coupling analysis and numerical analysis that
na\"ive fermions break the $Z^4_d$ symmetries
but overlap fermions do not. The reason for this difference
is a certain $Z_2$ symmetry that is broken
by na\"ive fermions. 

Fermions play a dynamical role and we opted to use
the hybrid Monte Carlo algorithm for generating gauge field
configurations. Since we are on a single site lattice, the
size of the fermion matrix is $4(N^2-1)\times 4(N^2-1)$ and
we compute the fermionic force exactly. We perform a full 
inversion of the overlap Dirac matrix using an
exact diagonalization of its kernel,
the Wilson-Dirac matrix.
We present the details of the underlying
hybrid Monte Carlo algorithm whose computational
cost scales like $N^6$.
We will show in this paper that one can use this algorithm to
compute physics quantities like the chiral condensate.
Detailed physics results will be presented
in~\cite{hnprep}.

\section{The single site model}

The partition function on a single site lattice 
is given by
\be
Z = \int [dU] e^{-S}=\int [dU] e^{-S_g - S_f},
\ee
where
\be
S_g = -bN \sum_{\mu\ne\nu=1}^4
\Tr  \left [ U_\mu U_\nu U_\mu^\dagger U_\nu^\dagger\right],
\label{gaction}
\ee
is the standard Wilson gauge action. With $f$ flavors of Dirac
fermions,
the contribution to the action from fermions is
\be
S_{n,o} = -f \log \det H_{n,o},\label{sno}
\ee
where $\gamma_5 H_{n,o}$ is the na\"ive (n) or
overlap (o) Dirac operator.
The $N\times N$ matrices
$U_\mu$; $\mu=1,2,3,4$ belong to $SU(N)$.
The lattice gauge coupling constant is $b=\frac{1}{g^2N}$.
The fermions are in the adjoint representation and they 
couple to the gauge fields, $V_\mu$, given by
\be 
V_\mu^{ab} = \frac{1}{2}\Tr \left [ T^a U_\mu T^b U^\dagger_\mu\right],
\ee 
where
$T^a$, $a=1,\cdots, (N^2-1)$, are traceless hermitian matrices 
that generate the $su(N)$ lie algebra.  
We have chosen the generators, $T^a$, to satisfy
\be
\Tr T^a T^b = 2 \delta^{ab}.
\ee
The fully anti-symmetric structure constants are defined using
\be
[T^a, T^b] = \sum_c i f^{ab}_c T^c.\label{structure}
\ee

Both $H_n$ and $H_o$ are $4(N^2-1)\times 4(N^2-1)$ hermitian matrices
and correspond 
to na\"ive Dirac fermions
and overlap Dirac fermions respectively.
The determinant of $H_{n,o}$ is positive definite and therefore the
logarithm is well defined. Furthermore,
we will show in the following two subsections that there exists a
hermitian matrix $\Sigma$ such that
\be
\Sigma H_{n,o} \Sigma = H^*_{n,o},\label{hadjiden}
\ee
which implies that all eigenvalues of $H_{n,o}$ are doubly degenerate
reflecting the adjoint nature of the fermions.
In addition, both na\"ive and overlap fermions obey chiral symmetry
and therefore the eigenvalues of $H_{n,o}$ will come in $\pm$ pairs.
Therefore the factor, $f$, in front of $S^f$ can be an integer (single Dirac
flavor) 
or half-integer (single Majorana flavor) for all values of
$N$.\footnote{Note that one should have written $\gamma_5H_{n,o}$ in
(\ref{sno}) but this is the same as writing $H_{n,o}$ as long as $f$ is
an integer multiple of $\frac{1}{2}$.} Since $f$ appears only as 
a multiplicative factor in the fermion action, we may take 
the number of fermions flavors, $f$, to
be any real number and we will do so in this paper. 

\subsection{Na\"ive fermions}
The matrix $H_n$ is given by
\be
H_n =  \pmatrix{\mu & C \cr C^\dagger & -\mu}
\label{faction}
\ee 
where $\mu$ is the fermion mass. 
The chiral matrix, $C$, is
\be
C = \frac{1}{2}\sum_\mu \sigma_\mu \left(V_\mu - V_\mu^t\right);
\label{vadj}
\ee
\be
\sigma_1=\pmatrix{0 & 1 \cr 1 & 0\cr};\ \ 
\sigma_2=\pmatrix{0 & -i \cr i & 0\cr};\ \ 
\sigma_3=\pmatrix{1 & 0 \cr 0 & -1\cr};\ \ 
\sigma_4=\pmatrix{i & 0 \cr 0 & i\cr}.
\ee 
Note that
\be
\sigma_2 \sigma_\mu^* \sigma_2 = -\sigma_\mu. \label{ccsigma}
\ee
Since
$V_\mu$ are real matrices, it follows from
(\ref{ccsigma}) that
\be
C^* = -\sigma_2 C \sigma_2. \label{cidenadj}
\ee
If we define
\be
\Sigma = \pmatrix{\sigma_2 & 0 \cr 0 & -\sigma_2\cr};\ \ \ 
\Sigma^\dagger = \Sigma;\ \ \ \Sigma^2=1,\label{sigdef}
\ee
then (\ref{hadjiden}) follows from (\ref{cidenadj}).

\subsection {Overlap fermions}\label{overlap}

The hermitian Wilson Dirac operator 
is
given by
\bea
H &=& \pmatrix{ 4 - m -\frac{1}{2}\sum_\mu \left( V_\mu + V_\mu^t\right)
& \frac{1}{2}\sum_\mu \sigma_\mu \left(V_\mu - V_\mu^t\right) \cr
-\frac{1}{2}\sum_\mu \sigma^\dagger_\mu \left(V_\mu - V_\mu^t\right) &
-4 + m +\frac{1}{2}\sum_\mu \left( V_\mu + V_\mu^t\right)\cr}\cr
&=& (4-m)\gamma_5
- \sum_\mu \left ( w_\mu V_\mu + w_\mu^\dagger V_\mu^t\right)
\label{wilson}
\eea
with $m$ being the Wilson mass parameter and
\be
w_\mu = \frac{1}{2}
\pmatrix { 1 & -\sigma_\mu \cr \sigma_\mu^\dagger & -1\cr}.
\ee
Using the definition of $\Sigma$ from (\ref{sigdef}), we have
\be
\Sigma w_\mu \Sigma = w_\mu^*;\ \ \ \Sigma\gamma_5\Sigma=\gamma_5,
\label{wsymm}
\ee
and therefore it follows from (\ref{wilson}) that
\be
\Sigma H \Sigma = H^*.\label{hsymm}
\ee

The hermitian 
massive overlap Dirac matrix, $H_0$, is defined by
\be
H_o = \frac{1}{2}\left [ \left( 1 + \mu \right)\gamma_5 +
\left(1-\mu\right)\epsilon(H)\right],\label{hover}
\ee
where $\mu\in[0,1]$ is the bare mass.
(\ref{hadjiden}) for overlap fermions follows from (\ref{hsymm}).

\subsection{Fermion boundary conditions}

We have assumed periodic boundary conditions for fermions in the
previous two subsections. Other choices of boundary conditions 
that do not generate a U(1) anomaly amount
to replacing $V_\mu$ by $V_\mu e^{i\frac{2\pi k_\mu}{N}}$ with integer
valued $k_\mu$~\cite{Poppitz:2008hr}.
Physical results are expected to depend on the choice of boundary
conditions. This is in contrast to the case of large $N$ gauge
theories coupled to fundamental fermions. In that case, the
$e^{i\frac{2\pi k_\mu}{N}}$ factor can be absorbed by a change of gauge fields
that only changes the Polyakov loop and not the action. If the
$Z_N$ symmetries are not broken as is the case in the confined
phase, this change will not affect
physical results.

\section{Weak coupling analysis}\label{wca}

Our aim is to study whether the $Z_N^4$ symmetries are broken in the
weak coupling limit. We 
follow~\cite{Bhanot:1982sh} and perform the weak coupling
analysis by decomposing $U_\mu$ according to
\be U_\mu = e^{ia_\mu} D_\mu e^{-ia_\mu};\ \ \ \ 
D_\mu^{ij}=e^{i\theta_\mu^i}\delta_{ij}. 
\ee
Keeping $\theta_\mu^i$ fixed, we expand in powers of
$a_\mu$.
The lowest contribution to $S_g$ comes from the quadratic term in
$a_\mu$~\cite{Bhanot:1982sh} and the lowest contribution
to $S_f$ comes from setting $a_\mu=0$.
Each $V_\mu$ has $\frac{N(N-1)}{2}$
two by two blocks of the form
\be
\pmatrix{
\cos(\theta_\mu^i-\theta_\mu^j) &   
\sin(\theta_\mu^i-\theta_\mu^j)   \cr
-\sin(\theta_\mu^i-\theta_\mu^j)   &
\cos(\theta_\mu^i-\theta_\mu^j)   \cr}
\ee 
with $1\le i < j \le N$. The remaining $(N-1)\times (N-1)$ matrix is a
unit matrix.
Therefore, the gauge field effectively has $(N-1)$ zero momentum modes
and $N(N-1)$ non-zero momentum modes of the form
$e^{i(\theta_\mu^i-\theta_\mu^j)}$ with $1\le i \ne j \le N$.

The computation of the fermion determinant
reduces to a free field calculation at this order
and the result is
\be 
S_{n,o} = -4f \sum_{i\ne j} \ln \lambda_{n,o}( \theta^i-\theta^j+\phi)-4(N-1)f\ln\lambda_{n,o}(\phi)
\ee 
where $e^{i\phi_\mu}$,
$\phi_\mu = \frac{2\pi k_\mu}{N}$, is the phase associated with the boundary
condition in the $\mu$ direction.
The eigenvalues, $\pm\lambda(p)$, are two fold degenerate and
the explicit expressions are
\be
\lambda_n(p) = \sqrt{\mu^2 + \sum_\mu \sin^2 p_\mu}\label{lamnaive}
\ee
for naive fermions
and 
\be
\lambda_o(p) =  \sqrt{\frac{1+\mu^2}{2} + \frac{1-\mu^2}{2} 
\frac{2\sum_\mu\sin^2\frac{p_\mu}{2} -m}
{\sqrt{
\left(2\sum_\mu\sin^2\frac{p_\mu}{2} -m\right)^2 + \sum_\mu \sin^2 p_\mu}}}\label{lamoverlap}
\ee
for overlap fermions.
The complete result from fermions and gauge fields is
\be
S =  \sum_{i\ne j} \left\{
\ln \left[\sum_\mu \sin^2 \frac{1}{2}\left(\theta_\mu^i-\theta_\mu^j\right) \right]
- 4f
\ln \lambda_{n,o}( \theta^i-\theta^j+\phi)\right\} -4(N-1)f\ln\lambda_{n,o}(\phi).
\label{pertact}
\ee

Independent of the actual values of $\theta_\mu^i$, the fermion
eigenvalues will have $(N-1)$ zero modes with periodic
boundary conditons when the mass is set to zero.
If all the $\theta_\mu^i$ are different for each $\mu$, then the
fermions should not have exact zero modes when we set
$p_\mu$ equal to $\left(\theta_\mu^i-\theta_\mu^j\right)$ with $i\ne j$. If the fermion spectrum
has more than $(N-1)$ zero modes, we will refer to the extra ones as doubler zero
modes.

In order to find the minimum of $S$, we consider the Hamiltonian
\be
H = \frac{1}{2}\sum_{\mu,i} \left(\pi_\mu^i\right)^2 + \beta S.
\ee
For large $\beta$, the Boltzmann measure $e^{-H}$ will be dominated
by the minimum of $S$. We can perform a HMC update of the $\pi,\theta$
system to find this minimum. The equations of motion for our $H$ are
\be
\dot \theta_\mu^i = \pi_\mu^i\label{tdot}
\ee
and
\be
\dot \pi_\mu^i = -\beta \frac{\partial S}{\partial \theta_\mu^i}
= -\beta\left(F^i_g -2f F^i_{n,o}\right).\label{pidot}
\ee
The gauge contribution to the force is
\be
F^i_g = \sum_{j\ne i} \frac{\sin\left(\theta_\mu^i - \theta_\mu^j\right)}
{\left[\sum_\nu \sin^2
  \frac{1}{2}\left(\theta_\nu^i-\theta_\nu^j\right) \right]}.
\label{fg}
\ee
The na\"ive fermion contribution to the force is
\be
F^i_n = \sum_{a=\pm}\sum_{j\ne i} \frac{\sin
\left[2p_{\mu a}^{ij}\right]}
{\mu^2 + c^{ij}_a},
\label{ffn}
\ee
and the overlap fermion contribution to the force is
\be
F^i_o = \frac{1-\mu^2}{2}\sum_{a=\pm}\sum_{j\ne i}\frac{
\sin p_{\mu a}^{ij} c^{ij}_a
- \frac{1}{2}\sin \left[2p_{\mu a}^{ij}\right]
b^{ij}_a}
{
\frac{1+\mu^2}{2}\left[d^{ij}_a\right]^{3/2}
+\frac{1-\mu^2}{2}b^{ij}_a d^{ij}_a},
\label{ffo}
\ee
where
\bea
c^{ij}_\pm &=& \sum_\nu \sin^2 p_{\nu \pm}^{ij}\cr
b^{ij}_\pm &=& 2 \sum_\nu \sin^2 \frac{p_{\nu \pm}^{ij}}{2}  - m\cr
d^{ij}_\pm &=& \left[ b^{ij}_\pm\right]^2 + c^{ij}_\pm\cr
p_{\mu\pm}^{ij} &=& \theta_\mu^i - \theta_\mu^j\pm \phi_\mu.
\eea

It is clear from (\ref{pidot}--\ref{ffo}) that 
\be
\sum_i \dot\pi^i_\mu = 0.
\ee
If we start with 
\be
\sum_i \pi^i_\mu = 0,
\ee
then it will remain zero. Then it follows from (\ref{tdot}) that
\be
\sum_i\dot\theta^i_\mu = 0.
\ee
If we start with
\be
\sum_i \theta^i_\mu =0,
\ee
then it will remain zero. These are just the conditions for remaining
in SU(N).

A choice for
the order parameters associated with the $Z_N^4$ symmetries is~\cite{Bhanot:1982sh}
\be
P_\mu = \frac{1}{2} \left( 1 - \frac{1}{N^2}|\Tr U_\mu|^2\right)
=\frac{1}{N^2}\sum_{i,j}
\sin^2 \frac{1}{2}\left(\theta_\mu^i-
\theta_\mu^j\right) 
\ee
If $P_\mu=\frac{1}{2}$, then the $Z_N$ symmetry in that direction is
not broken. 
If $\theta_\mu^i$ are uniformly distributed in a width $\alpha\le
2\pi$, then
\be
\lim_{N\to\infty} P_\mu= \frac{1}{2}\left[ 1-
  \left(\frac{2}{\alpha}\sin\frac{\alpha}{2}\right)^2\right].\label{palpha}
\ee

\subsection{Na\"ive fermions break the $Z_N^4$ symmetries}

\begin{figure}
\centerline{\includegraphics[width=0.8\textwidth]{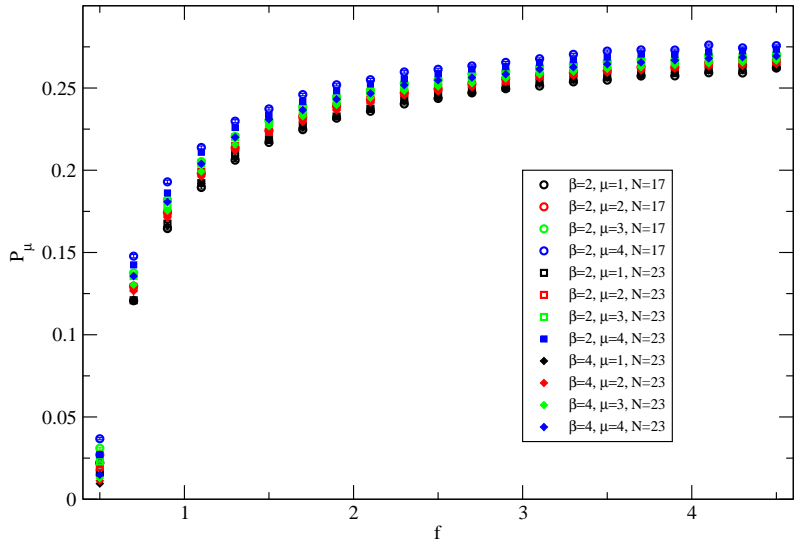}}
\caption{Na\"ive fermions:
Plot of $P_\mu$ as a function of $f$ for three different
 choices of $(\beta ,N)$ and
$\mu=0.01$.}
\label{fig1}
\end{figure}

\begin{figure}
\centerline{\includegraphics[width=0.8\textwidth]{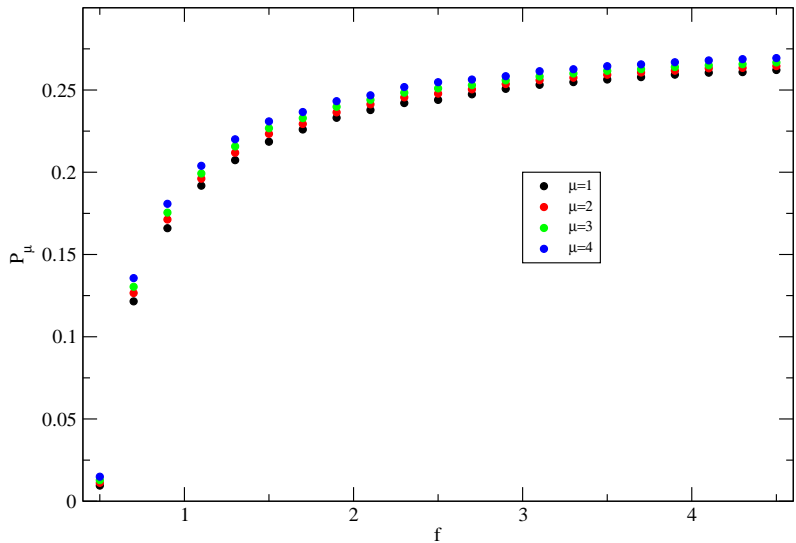}}
\caption{Na\"ive fermions:
Plot of $P_\mu$ as a function of $f$ at $N=23$, $\beta=4$, and
$\mu=0.01$.}
\label{fig2}
\end{figure}

We assume periodic boundary conditions and set $\phi_\mu=0$ in
(\ref{pertact}). We pick one value of $N$ and $\beta$ and calculate $P_\mu$
as a function of $f$. In order to clearly see symmetry breaking we
use rotational symmetry on the lattice and
choose to label our directions such 
that $P_1< P_2 < P_3 < P_4$ for each configuration in our
thermalized ensemble. We set $\mu=0.01$ to avoid potential
singularities that could occur for the massless case.
The plot for three different choices of $(\beta,N)$ 
is shown in Fig.~\ref{fig1}. The value is well below $\frac{1}{2}$
and it seems to approach a value for large $f$ that is consistent
with $\alpha=\pi$ in (\ref{palpha}). Since all three choices
for $(\beta,N)$ give consistent values, we have plotted the
results for $\beta=4$ and $N=23$ in Fig.~\ref{fig2}.
The small deviations one sees between the four different
$P_\mu$ are simply due to our ordering scheme combined
with finite $N$ effects. 

\begin{figure}
\centerline{\includegraphics[width=0.8\textwidth]{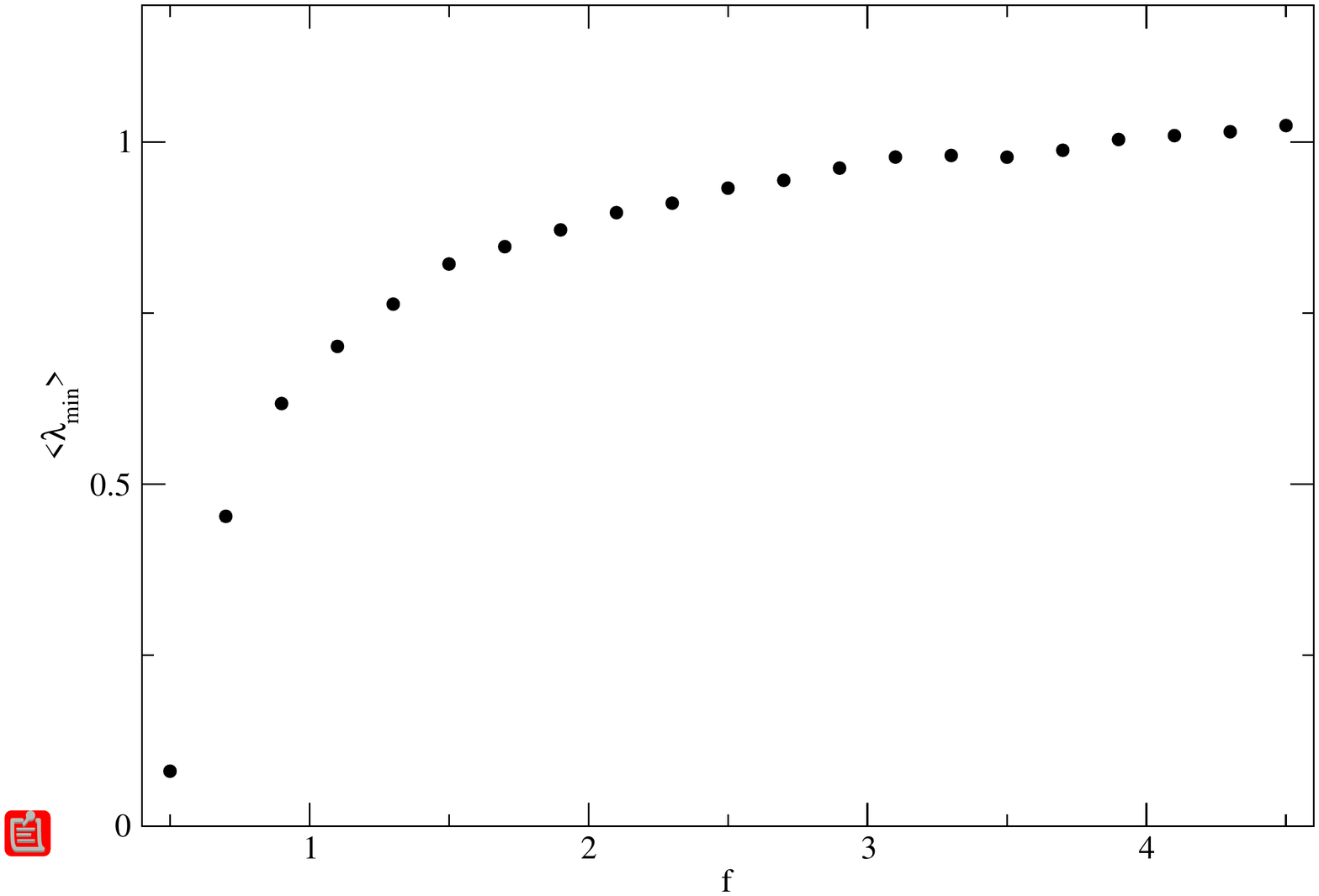}}
\caption{Na\"ive fermions:
Plot of 
smallest eigenvalue,
$\lambda(p)$ with $p_\mu=\left(\theta_\mu^i-\theta_\mu^j\right)$
and $i\ne j$, for
$N=23$, $\beta=4$, and
$\mu=0.01$.}
\label{fig3}
\end{figure}

With $N=23$, we expect $22$ exact zero modes 
for $\lambda(p)$ as explained in
section~\ref{wca}. We plot the average of the smallest 
eigenvalue,
$\lambda(p)$ with $p_\mu=\left(\theta_\mu^i-\theta_\mu^j\right)$ 
and $i\ne j$, in Fig.~\ref{fig3}. 
It is clear that this eigenvalue is non-zero for
all
values of $f$ indicating that there are no doubler zero modes.
The reason for the breaking of the $Z_N^4$ symmetries can be
understood by looking at the total action obtained from
(\ref{lamnaive}) and (\ref{pertact}):
\be S =  \sum_{i\ne j} 
\ln \left[\sum_\mu
\sin^2 \frac{1}{2}\left(\theta_\mu^i-
\theta_\mu^j\right) \right]
-2f \sum_{i\ne j} \ln \left[
\mu^2 + \sum_\mu \sin^2 (\theta^i_\mu-\theta^j_\mu)\right].
\ee
The fermionic contribution cannot separate 
$\theta_\mu^i=\theta_\mu^j$ 
from $\theta_\mu^i=\theta_\mu^j+\pi$  
implying that the fermion
contribution alone will result in a distribution of eigenvalues
restricted to a width of $\pi$.

\subsection{Overlap fermions do not break the $Z_N^4$ symmetries}

\begin{figure}
\centerline{\includegraphics[width=0.8\textwidth]{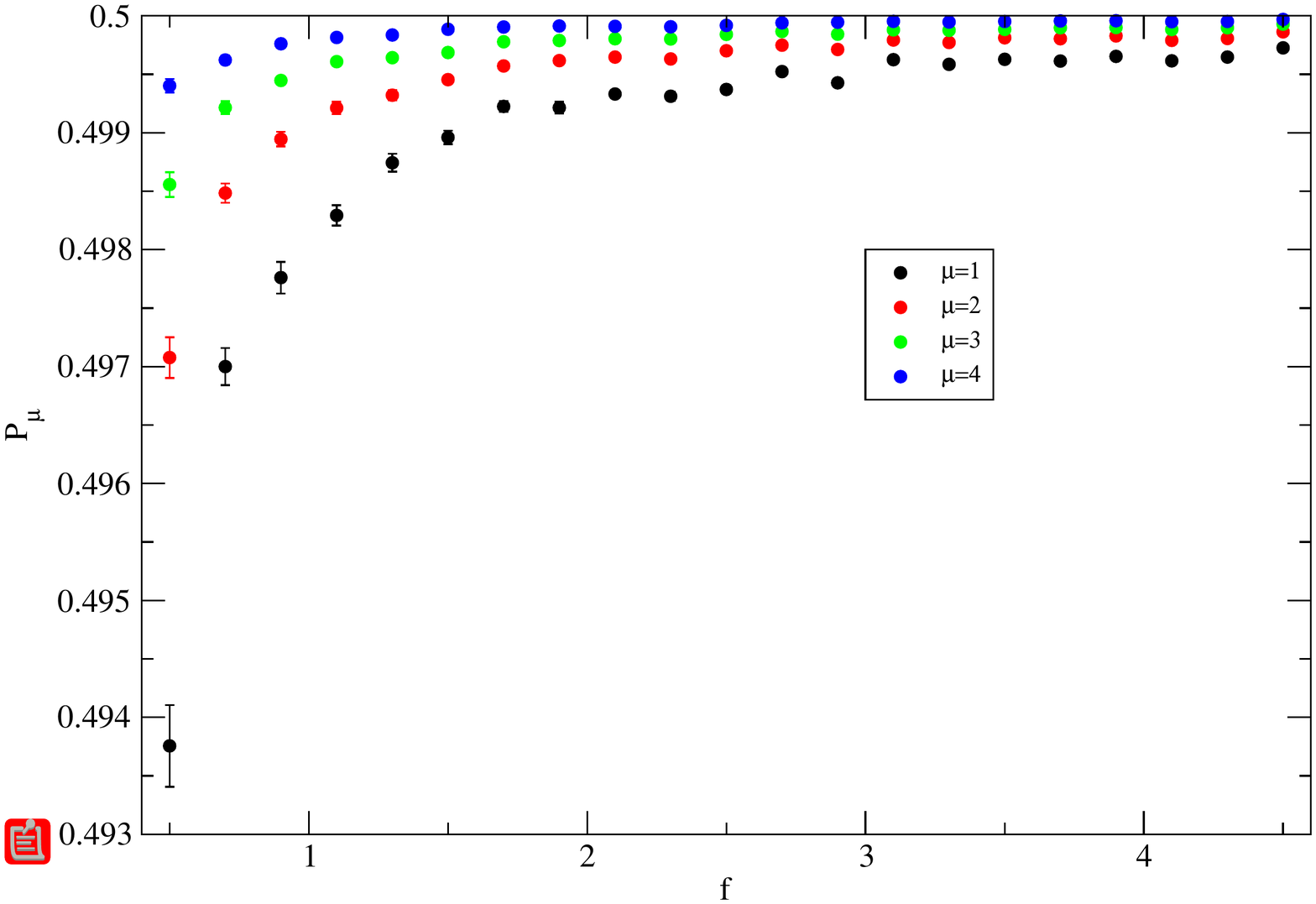}}
\caption{Overlap fermions:
Plot of $P_\mu$ as a function of $f$ at $N=23$, $\beta=1$, $m=5$, and
$\mu=0.01$.}
\label{fig4}
\end{figure}

Contrary to na\"ive fermions, overlap fermions do separate 
$\theta_\mu^i=\theta_\mu^j$
from $\theta_\mu^i=\theta_\mu^j+\pi$ 
as is evident from the presence of the Wilson factors in
(\ref{lamoverlap}).
Therefore, we do not expect overlap fermions to break the
$Z_N^4$ symmetries. A plot of $P_\mu$ for several values of $f$
at $N=23$, $\beta=1$, and $m=5$ with $\mu=0.01$ in Fig.~\ref{fig4}
shows this to be the case. The very small deviation close to
$f=\frac{1}{2}$
is a consequence of finite $N$ effects. Since we do not expect
to go beyond $N=23$ in the full simulation of the model, this
plot will serve as a guide to what one can expect in a full
simulation. 

\begin{figure}
\centerline{\includegraphics[width=0.8\textwidth]{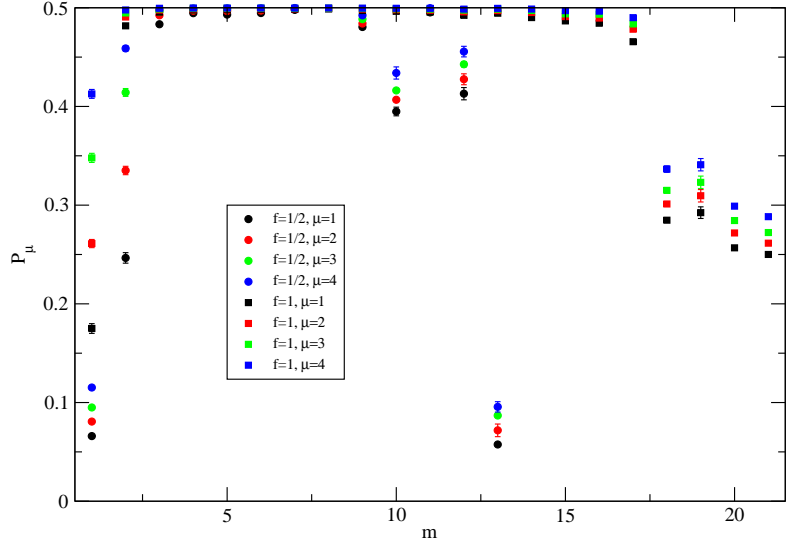}}
\caption{Overlap fermions:
Plot of $P_\mu$ as a function of $m$ at $N=23$, $\beta=1$, and
$\mu=0.01$ for two different values of $f$.}
\label{fig5}
\end{figure}

\begin{figure}
\centerline{\includegraphics[width=0.8\textwidth]{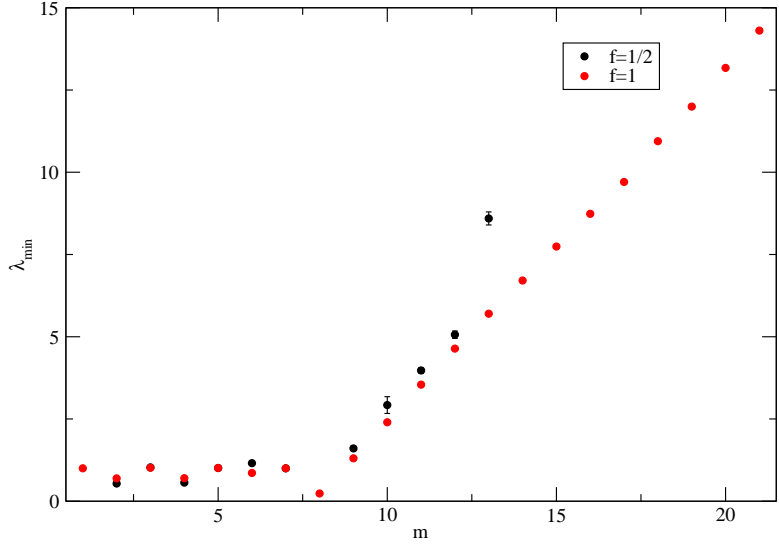}}
\caption{Overlap fermions:
Plot of the lowest positive eigenvalue of $H$
as a function of $m$ at $N=23$, $\beta=1$, and
$\mu=0.01$ for two different values of $f$.}
\label{fig6}
\end{figure}

\begin{figure}
\centerline{\includegraphics[width=0.8\textwidth]{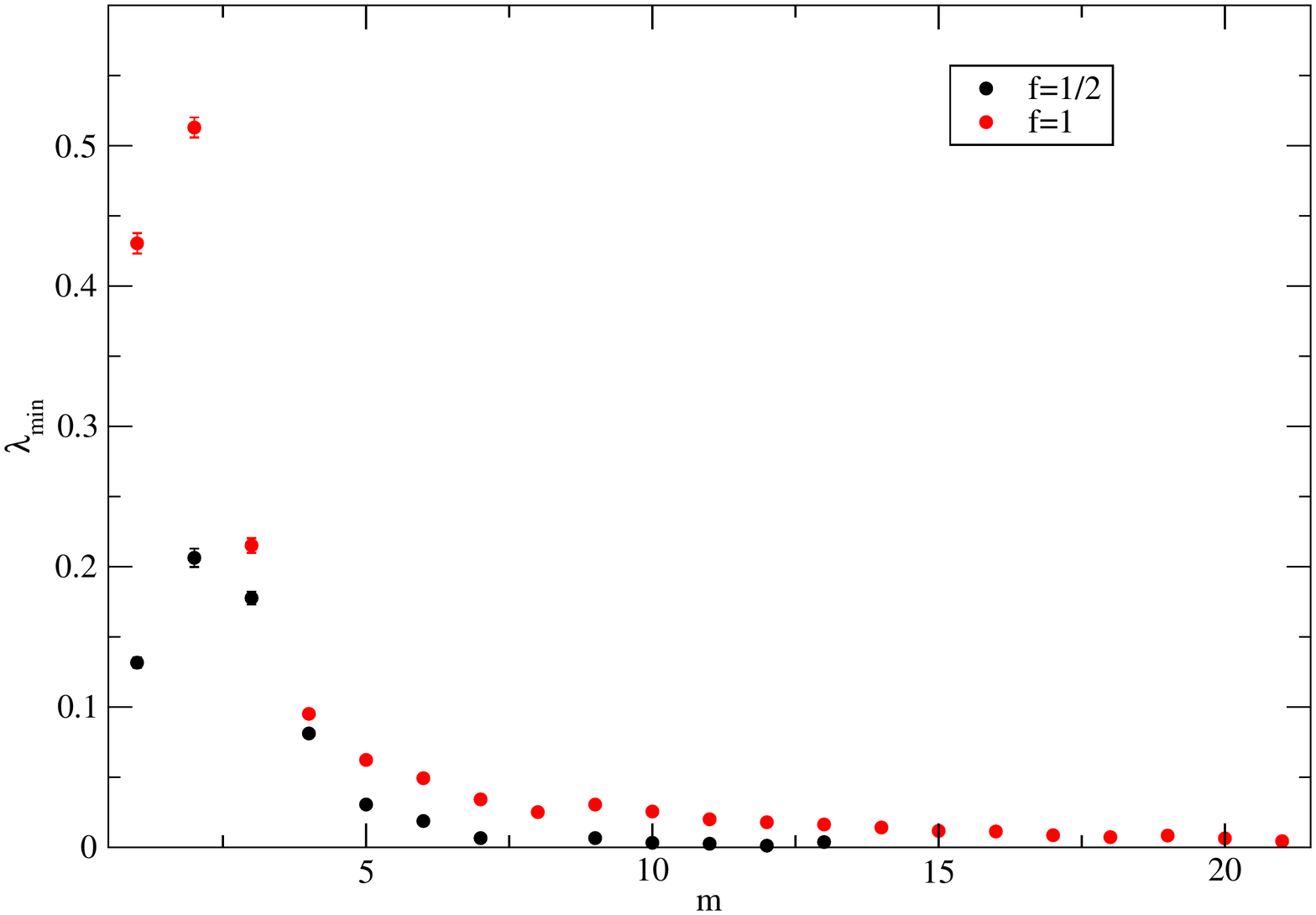}}
\caption{Overlap fermions:
Plot of  
smallest eigenvalue,
$\lambda(p)$ with $p_\mu=\left(\theta_\mu^i-\theta_\mu^j\right)$ 
and $i\ne j$, for
$N=23$, $\beta=1$, and
$\mu=0.01$.}
\label{fig7}
\end{figure}

Since there are no doubler zero modes, we do not have any
restriction on the values for the Wilson mass, $m$, used in the
Wilson-Dirac kernel as described in section~\ref{overlap}.
But, we cannot make it arbitrarily large since one can
see by a direct computation that the large $m$ limit
of overlap fermions is na\"ive fermions~\cite{Narayanan:1994gw}.
A plot of $P_\mu$ as a function of $m$ is shown for $f=\frac{1}{2}$
and $f=1$ in Fig.~\ref{fig5}. It indicates that $3\le m \le 8$ is an
appropriate range of values of $m$ where the $Z_N^4$ symmetries
are not broken for $f=\frac{1}{2}$ and that range only gets bigger
as $f$ increases. A plot of the lowest positive eigenvalue of $H_w$ in Fig.~\ref{fig6}
and the smallest eigenvalue, $\lambda_o(p)$ with
$p_\mu=\left(\theta_\mu^i-\theta_\mu^j\right)$, $i\ne j$, in Fig.~\ref{fig7}
shows that
there are no doubler zero modes in this range of $m$.
We can use this range of $m$ for our full numerical simulation
with overlap fermions.
One cannot take the Wilson mass parameter to zero in
the weak coupling limit. This is intimately tied to the
fact that the $Z_N$ symmetries are not broken 
and $U_\mu\ne 1$ in the weak coupling limit.

\subsection{Effect of boundary conditions}

\begin{figure}
\centerline{\includegraphics[width=0.8\textwidth]{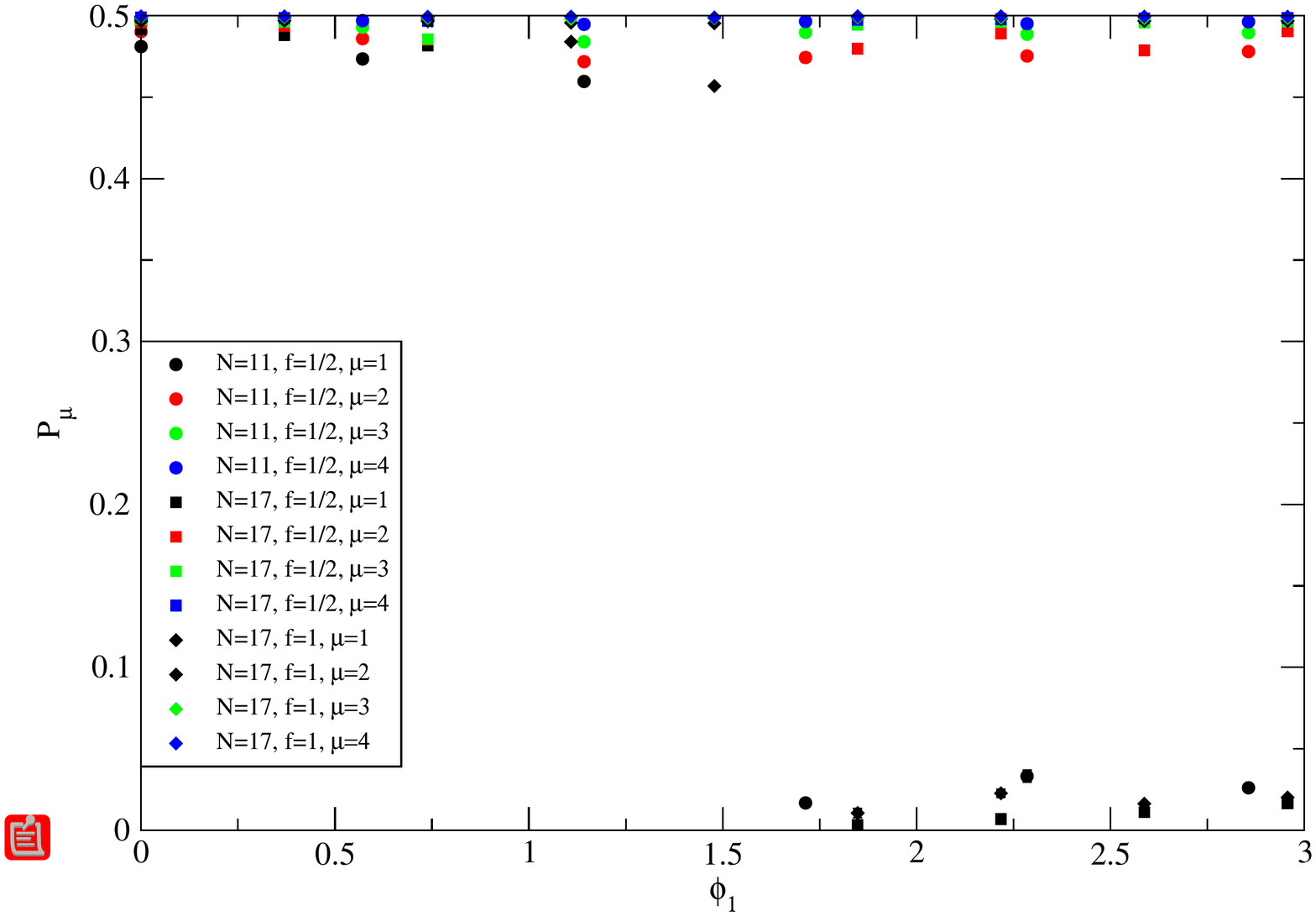}}
\caption{Overlap fermions:
Plot of $P_\mu$ as a function of $\phi_1$ for three different values
of $(f,N)$ with $\beta=1$ and
$\mu=0.01$.}
\label{fig8}
\end{figure}

In order to study the effect of boundary conditions on $Z_N^4$
symmetry breaking, we focus on $f=\frac{1}{2}$ and $f=1$.
We set $\phi_\mu=0$ for $\mu=2,3,4$ and varied $\phi_1$
by setting it to equal to $\frac{2\pi k}{N}$ with $k$ an
integer in the range $0\le k < N/2$. The plot of $P_\mu$
as a function of $\phi_1$ is shown in Fig.~\ref{fig8}. We see
that the $Z_N$ symmetry in the $\mu=1$ direction is broken
if $\phi_1 > \frac{\pi}{2}$. This seems to be the case in
the limit of large $N$ and seems to be roughly independent
of $f$. Furthermore, the $Z_N$ symmetries in the other three
directions with periodic boundary conditions are not broken.
This result could help us force feed momentum for quarks
in the adjoint representation. In order to pursue this, we
need to study the effect of $\phi_1$ on the chiral condensate
and see if chiral symmetry is restored when
the $Z_N$ symmetry is broken and if the chiral condensate
is independent of the value of $\phi_1$ when the $Z_N$ symmetry
is not broken.

\section{Hybrid Monte Carlo algorithm}

One could use the standard Metropolis algorithm to study the single
site model but the computational cost would scale like $N^8$~\cite{Bringoltz:2009kb}.
We will use the Hybrid Monte Carlo 
algorithm~\cite{DeGrand:2006zz} to numerically
study
the model on a single site lattice.  
Fermion matrix inversions
 dominate the  computational cost of this algorithm
and will therefore scale like $N^6$.
We set $H_\mu^{ij}$
as
momentum variables conjugate to $U_\mu^{ij}$ with the condition
that $H_\mu$ is a traceless hermitian matrix. 
Therefore, we can write
$H_\mu$ as
\be 
H_\mu = \sum_a H_{\mu a} T^a.\label{hmua}
\ee

The Hamiltonian
for the Hybrid Monte Carlo algorithm is
\be
{\cal H} = \frac{1}{2}\sum_\mu \Tr H_\mu^2 + S.
\ee
Since the algorithm has not been used in the past for gauge theories
with adjoint fermions on a single site lattice and since we do
not use the conventional pseudo-fermion algorithm for dealing
with the fermionic determinant, we present the necessary details
in this section.

The equations of motion for $U_\mu$ are
\be
\frac{ d U_\mu}{d\tau} = i H_\mu U_\mu.\label{ueqn}
\ee
Setting $\frac{d {\cal H}}{d\tau}=0$ 
results in
\be
\sum_{\mu} \Tr \left [ H_\mu \frac{ dH_\mu}{d\tau} \right]+
\frac{ dS_g}{d\tau} + \frac{ dS_{n,o}}{d\tau}=0,
\ee
where we can write
\bea
\frac{ dS_g}{d\tau} &=& \sum_{\mu}  \Tr 
\left [ H_\mu D^g_\mu\right]\cr
\frac{ dS_{n,o}}{d\tau} &=& \sum_{\mu} \Tr 
\left [ H_\mu D^{n,o}_\mu\right].
\eea

A simple calculation results in
\be
D^g_\mu = -ibN \sum_{\nu} \left [
U_\mu U_\nu U_\mu^\dagger U_\nu^\dagger
+ U_\mu U_\nu^\dagger U_\mu^\dagger U_\nu
- U_\nu^\dagger U_\mu U_\nu U_\mu^\dagger
- U_\nu U_\mu U_\nu^\dagger U_\mu^\dagger  \right ]
\ee

The derivative of $V_\mu$ is
\be
\frac{ dV_\mu^{ab}}{d\tau} =
- \sum_{cd} H_{\mu c} f^{ac}_d V_\mu^{db}
=
- \sum_{d} \bar H^a_{\mu d} V_\mu^{db}
\ee
where
\be
\bar H^a_{\mu d}\equiv \sum_c H_{\mu c} f^{ac}_d,
\ee
and we have used (\ref{structure}) and (\ref{hmua}) .
Using the anti-symmetry of the structure constants, it follows
that $\bar H_\mu$ are anti-symmetric real matrices.
In analogy with (\ref{ueqn}), we can write
\be
\frac{ dV_\mu }{d\tau} = - \bar H_\mu V_\mu.\label{dvmu}
\ee

\subsection{Derivation of $D^n_\mu$ for na\"ive fermions:}

We start by noting that
\be
\frac{ dS_n}{d \tau} = -f\Tr \frac{1}{m^2 +CC^\dagger} 
\left( \frac{dC}{d\tau} C^\dagger + C\frac{dC^\dagger}{d\tau}
\right),\label{sfder}
\ee 
and
\be
\frac{d C}{d\tau} = - \frac{1}{2}\sum_\mu \sigma_\mu\left(\bar H_\mu V_\mu
+V_\mu^t \bar H_\mu\right).
\ee
Let us define
\be
M_{\mu\nu} = \frac{1}{4}\tr \left [ \frac{1}{m^2 + CC^\dagger} 
\sigma_\mu\sigma_\nu^\dagger \right]; \ \ \ 
M_{\mu\nu}^\dagger = M_{\nu\mu},\label{Mmunu}
\ee
where $\tr$ stands for trace only over the spin index.
Using
(\ref{ccsigma}) and (\ref{cidenadj}) we see that
$M_{\mu\nu}$ are real matrices.
We can rewrite (\ref{sfder}) as
\be
\frac{ dS_n}{d \tau}
= - f \sum_{\mu} \Tr \bar H_\mu \bar A_\mu,
\ee
where
\be
\bar A_\mu = 
\left [
V_\mu B_\mu + B_\mu V_\mu^t 
\right] - {\rm transpose},
\ee
are real anti-symmetric matrices
and
\be
B_\mu = \sum_{\nu} (V_\nu - V_\nu^t)M_{\mu\nu}.
\ee
This brings us down to the form we need, namely,
\be
\frac{ dS_n}{d \tau}
= \sum_\mu \Tr H_\mu D^n_\mu
\ee
where
\be
D^n_\mu 
= i\frac{f}{2}\sum_{ab}\bar A^a_{\mu b} [T^a,T^b].
\ee

\subsection{Derivation of $D^o_\mu$ for overlap fermions:}

Let the eigenvalues of the Wilson-Dirac operator, $H$, given in (\ref{wilson}) be
\bea
H |\lambda_i\rangle =& \lambda_i |\lambda_i\rangle;&\lambda_i > 0\cr
H |\omega_i\rangle =& \omega_i |\omega_i\rangle;& \omega_i < 0\cr
\langle\lambda_i|\lambda_j\rangle =\delta_{ij};&
\langle\omega_i|\omega_j\rangle =\delta_{ij};&
\langle\lambda_i|\omega_j\rangle =0.
\eea
The dimension of $H$ is $4(N^2-1)\times 4(N^2-1)$ and we will assume that
there are an equal number of positive and negative
eigenvalues.\footnote{This is a reasonable assumption since 
we can restrict ourselves to a fixed topological sector.}
If follows from (\ref{hsymm}) that all eigenvalues of
$H$ are doubly degenerate since the pair of
eigenvectors ($\Sigma |\lambda_i\rangle^*, |\lambda_i\rangle$)
have the same eigenvalue and the same is the case for the
pair ($\Sigma |\omega_i\rangle^*, |\omega_i\rangle$).

The fermion action is defined by
\be
S_o = -f \Tr \log H_o,
\ee
and
\be
\frac{dS_o}{d\tau} = -f \Tr \frac{1}{H_o} \frac {d H_o}{d\tau}.
\label{dsfo}
\ee
It follows from (\ref{hover}) that
\be
\frac {d H_o}{d\tau} = \frac{1-\mu}{2} \frac{d\epsilon(H)}{d\tau}.
\label{dho}
\ee

We can write
\be
\epsilon(H) = \sum_i |\lambda_i\rangle\langle\lambda_i|
- \sum_i |\omega_i\rangle\langle\omega_i|.
\ee
Therefore,
\be
\frac{d\epsilon(H)}{d\tau} = 
\sum_i \frac{d |\lambda_i\rangle}{d\tau}\langle\lambda_i|
+\sum_i |\lambda_i\rangle\frac{d \langle\lambda_i|}{d\tau}
- \sum_i \frac{d|\omega_i\rangle}{d\tau}\langle\omega_i|
- \sum_i |\omega_i\rangle\frac{d\langle\omega_i|}{d\tau},
\ee
and
\be
\frac{d|\lambda_i\rangle}{d\tau} =
\sum_{j\ne i} \frac 
{\langle \lambda_j | \frac{dH}{d\tau} | \lambda_i\rangle}
{\lambda_i - \lambda_j}|\lambda_j\rangle
+\sum_{j} \frac 
{\langle \omega_j | \frac{dH}{d\tau} | \lambda_i\rangle}
{\lambda_i - \omega_j}|\omega_j\rangle,
\ee
\be
\frac{d|\omega_i\rangle}{d\tau} =
\sum_{j} \frac 
{\langle \lambda_j | \frac{dH}{d\tau} | \omega_i\rangle}
{\omega_i - \lambda_j}|\lambda_j\rangle
+\sum_{j\ne i} \frac 
{\langle \omega_j | \frac{dH}{d\tau} | \omega_i\rangle}
{\omega_i - \omega_j}|\omega_j\rangle.
\ee
The above
equations imply
\bea
\frac{d\epsilon(H)}{d\tau} |\lambda_k\rangle &=&
2\sum_j \frac{|\omega_j\rangle\langle\omega_j|\frac{dH}{d\tau}
|\lambda_k\rangle}{\lambda_k-\omega_j},\cr
\frac{d\epsilon(H)}{d\tau} |\omega_k\rangle &=&
2\sum_j \frac{|\lambda_j\rangle\langle\lambda_j|\frac{dH}{d\tau}
|\omega_k\rangle}{\lambda_j-\omega_k}.
\eea
Inserting  (\ref{dho}) into (\ref{dsfo}) and using
the above equations, we get
\be
\frac{dS_o}{d\tau}
= f(1-\mu) \Tr \left[ \frac{dH}{d\tau} (G+G^\dagger)\right],
\label{dsfo1}
\ee
with
\be
G= \sum_{j,k} \frac{ |\lambda_j\rangle\langle\lambda_j|
\frac{1}{H_o} |\omega_k\rangle\langle\omega_k|}{\omega_k - \lambda_j}.
\ee
The symmetries in (\ref{hsymm}) and (\ref{hadjiden}) imply that
\be
\Sigma G \Sigma = G^*.\label{gsymm}
\ee
Equations (\ref{dsfo1}), (\ref{wilson})  and (\ref{dvmu}) can be
combined to give
\be
\left[ \frac{dS_o}{d\tau}\right] = 
f(1-\mu) \Tr \bar H_\mu \left [ V_\mu \left( G+G^\dagger\right)
\omega_\mu - \omega_\mu^\dagger \left (G + G^\dagger\right ) V_\mu^t
\right].\label{dsfo2}
\ee
Let
\be
\bar B_\mu = \tr (G+G^\dagger) \omega_\mu,
\ee
where $\tr$ stands for sum over the spin indices.
Using (\ref{wsymm}) and (\ref{gsymm}) we can see that
$B_\mu$ are real.
With the definition of $B_\mu$, (\ref{dsfo2}) becomes
\be
\frac{dS_o}{d\tau} = 
f(1-\mu) \Tr \bar H_\mu \left [ V_\mu B_\mu - B_\mu^t V_\mu^t
\right]
= f(1-\mu) \Tr \bar H_\mu \bar A_\mu,
\ee
where
\be
\bar A_\mu = 
V_\mu B_\mu - B^t_\mu V^t_\mu
\ee
are real anti-symmetric matrices.
The rest of the steps are identical to that of na\"ive fermions.

\section{Numerical results}

In this section we present numerical results from our Hybrid Monte
Carlo algorithm and show that they support the conclusions reached in
section~\ref{wca}. 

\begin{figure}
\centerline{\includegraphics[width=0.8\textwidth]{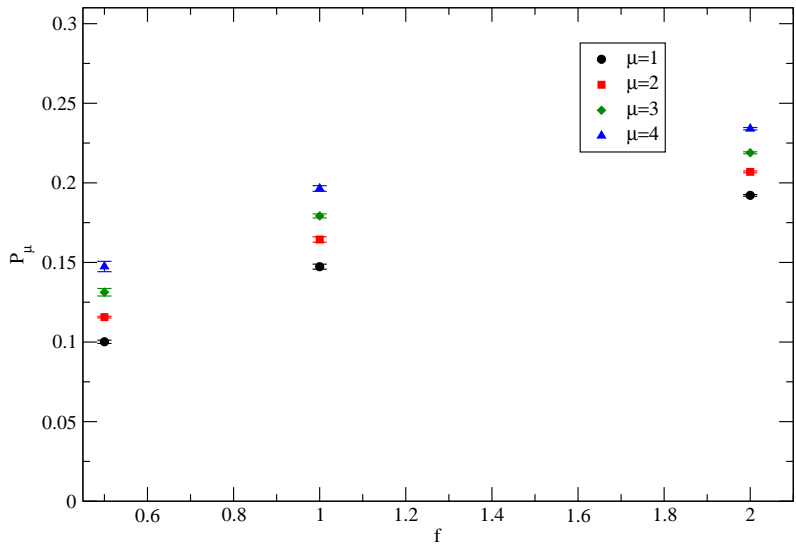}}
\caption{Na\"ive fermions:
Plot of $P_\mu$ as a function of $f$ for 
$N=11$, $b=5$ and $\mu=0.01$.}
\label{fig9}
\end{figure}

\begin{figure}
\centerline{\includegraphics[width=0.8\textwidth]{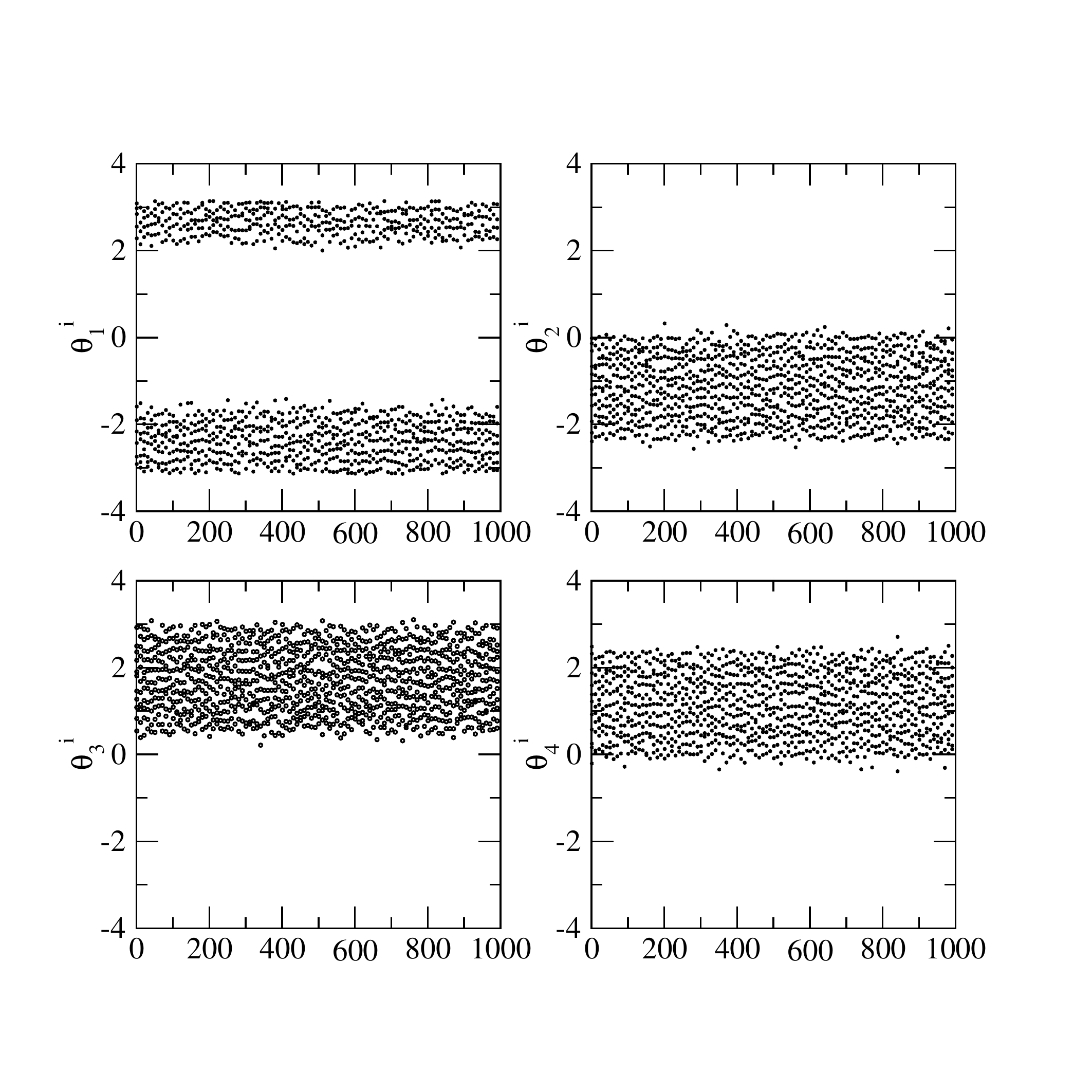}}
\caption{Na\"ive fermions:
Distribution of $\theta_\mu^i$ for
$N=11$, $b=5$ and $\mu=0.01$ at $f=2$.}
\label{fig10}
\end{figure}

Fig.~\ref{fig9} shows a plot of $P_\mu$ as a function of $f$ obtained using
na\"ive fermions. We have set $b=5$, $N=11$ and $\mu=0.01$.
Na\"ive fermions break the $Z_N^4$ symmetries as expected from the
weak coupling analysis. The values of $P_\mu$ in Fig.~\ref{fig9} are
consistent with the distribution of $\theta_\mu^i$ uniform with
a width less than $\pi$ and this is evident for $f=1$ in Fig.~\ref{fig10}.

\begin{figure}
\centerline{\includegraphics[width=0.8\textwidth]{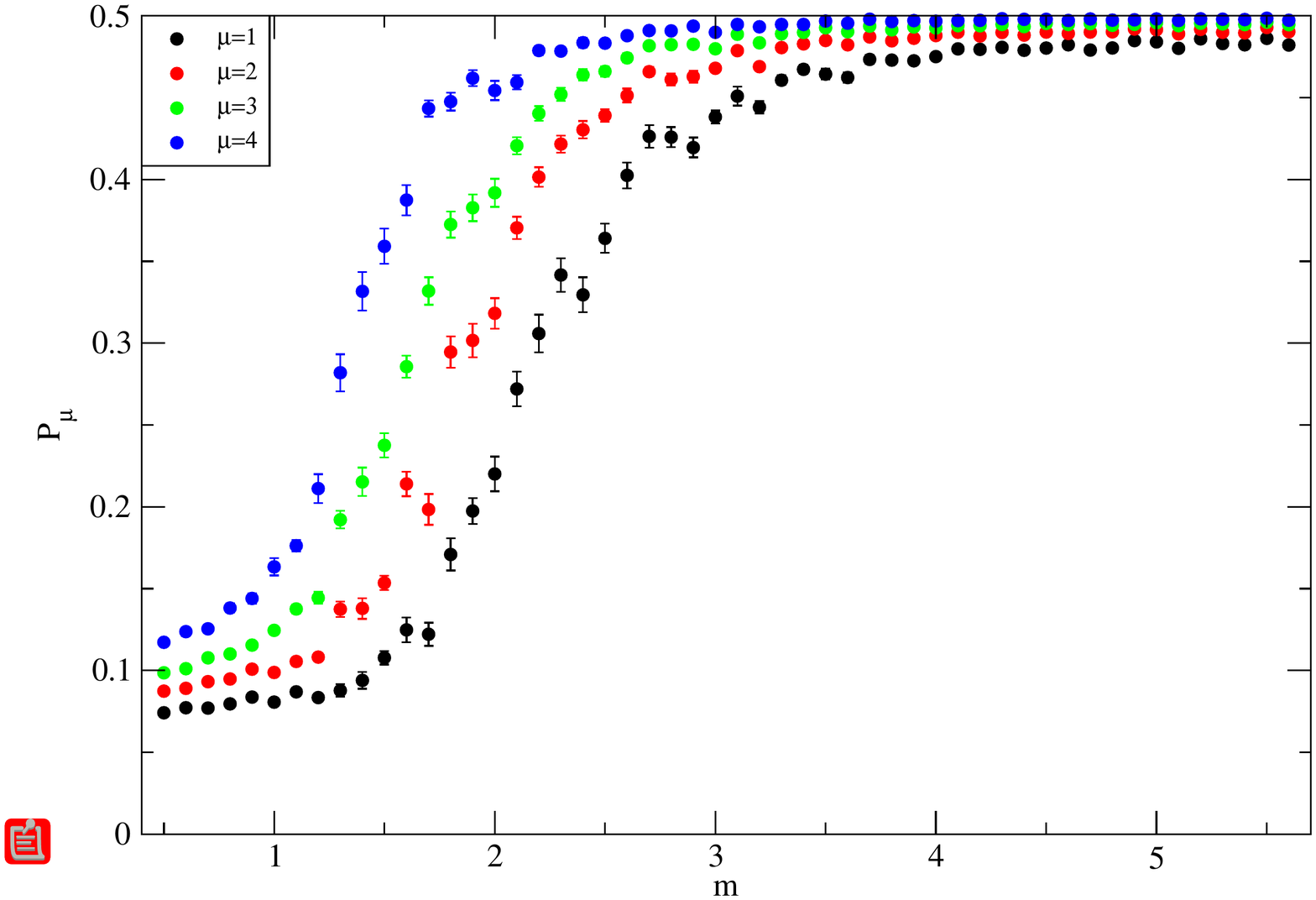}}
\caption{Overlap fermions:
Plot of $P_\mu$ as a function of $m$ for 
$N=11$, $b=7$, $f=\frac{1}{2}$ and $\mu=0.01$.}
\label{fig11}
\end{figure}

\begin{figure}
\centerline{\includegraphics[width=0.8\textwidth]{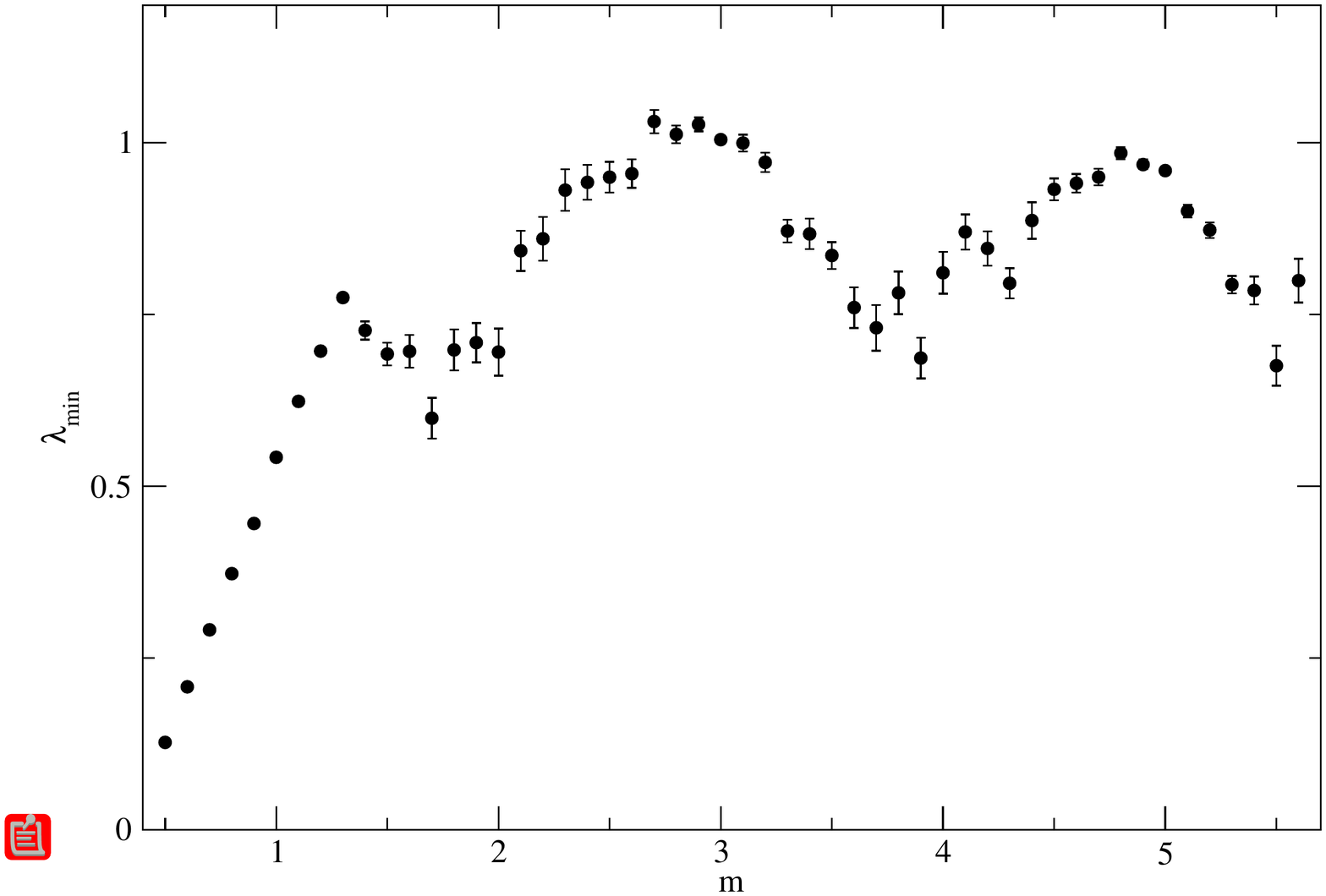}}
\caption{Overlap fermions:
Plot of the lowest eigenvalue of $H$ as a function of $m$ for 
$N=11$, $b=7$, $f=\frac{1}{2}$ and $\mu=0.01$.}
\label{fig12}
\end{figure}

Turning our attention to overlap fermions, we plot $P_\mu$ as a
function
of $m$ in Fig.\ref{fig11} for $b=7$, $N=11$, $f=\frac{1}{2}$ and $\mu=0.01$.
As expected the $Z_N^4$ symmetries are not broken for a wide range of
$m$. 
A plot
of the lowest eigenvalue of $H$ as a function of $m$ in Fig.~\ref{fig12}
shows that there are no doubler modes in the same wide range of $m$.

\begin{figure}
\centerline{\includegraphics[width=0.8\textwidth]{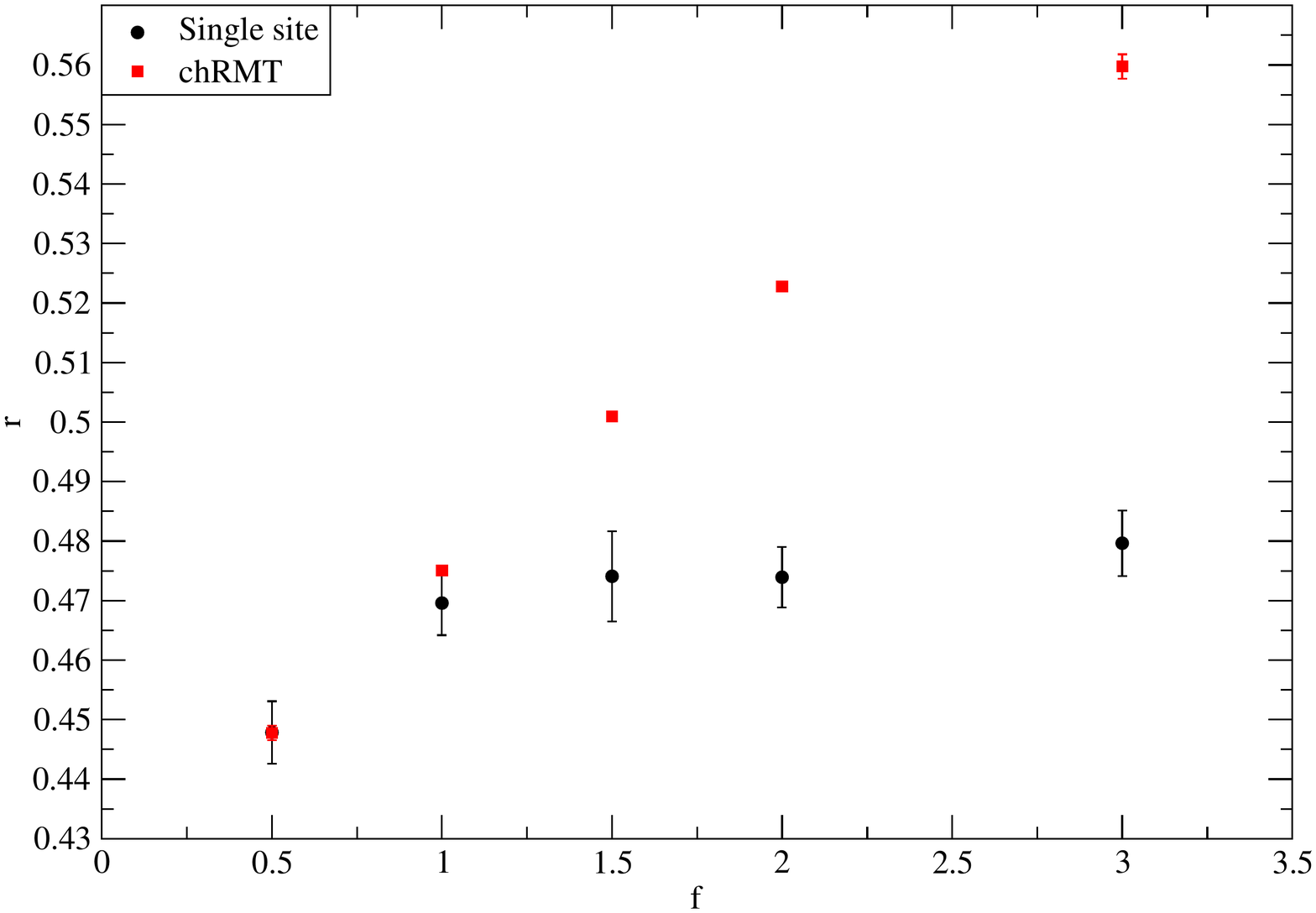}}
\caption{Overlap fermions:
Plot of the ratio, $r$, as a function of $f$ at
$N=11$, $b=5$, $m=5$ and $\mu=0.01$ is compared with chiral
random matrix theory predictions.}
\label{fig13}
\end{figure}

If chiral symmetry is broken, we expect the low lying eigenvalues
of the overlap Dirac matrix to scale like $\frac{1}{N^2-1}$.
Furthermore, we expect agreement with chiral random matrix
theory predictions for the distribution of the
low lying eigenvalues in accordance with the symplectic
ensemble~\cite{Damgaard:2000ah}.
Instead of using the analytical formulas in~\cite{Damgaard:2000ah} 
we found in convenient
to numerically generate the
eigenvalue distribution governed by the symplectic ensemble
using Hybrid Monte Carlo techniques
and compare with the numerical results obtained for this model.
The chiral Random Matrix theory ensemble for a symplectic
matrix, $C=\sum_\mu \sigma_\mu C_\mu$, is
\be
Z = \int [dC_\mu] e^{-\sum_\mu \sum_{ij} 
\left[ C_\mu^{ij}\right]^2}
\left[ \det H_{\rm rmt}\right]^f;\ \ \ \
H_{\rm rmt}=\pmatrix{ \mu & C \cr C^\dagger & -\mu \cr}
\ee
with $C_\mu$ being a real square matrix.
We expect one scale factor to relate the distributions of
the eigenvalues of $H_{\rm rmt}$ and $H_o$.
Since we want to eliminate the scale set by the chiral condensate,
we focus on 
\be
r= \left\langle \frac{\lambda_1}{\lambda_2} \right\rangle,
\ee
where $0 < \lambda_1 < \lambda_2$ are the first two non-degenerate
eigenvalues. 
Fig.~\ref{fig13} shows a plot of $r$ as a function of $f$ obtained at
$N=11$ and $b=5$ with $m=5$ and $\mu=0.01$.
The results are also compared with the numbers one gets from chiral
random matrix theory.
There is agreement with chiral random matrix theory for
$f=\frac{1}{2}$ and $f=1$ but not for $f=2$ or $f=3$.
The case of $f=\frac{1}{2}$ corresponds to the large $N$ limit of
super Yang-Mills
and this is one case where chiral symmetry assures supersymmetry.

One can try to understand the above behavior by considering the 
running of the coupling constant, $\alpha=\frac{1}{4\pi^2 b}$
with the scale set by the lattice spacing, $a$~\cite{Creutz:1984mg}.
Using the 
two-loop beta function~\cite{Caswell:1974gg,Jones:1974mm},
we have
\be
  a\frac{d\alpha}{da}=2\left(\frac{11}{3}-\frac{4}{3}f\right)\alpha^2+
2\left(\frac{34}{3}-\frac{32}{3}f\right)\alpha^3+\cdots,
\ee
where the first two coefficients listed are independent of the
the renormalization scheme. Asymptotic freedom is
lost when $f>\frac{11}{4}$.
The two-loop expression predicts an infrared fixed
point~\cite{Banks:1981nn} 
if $\frac{17}{16} < f < \frac{11}{4}$ with the fixed point
coupling becoming smaller as $f$ increases in this range.
Our current data (Fig.~\ref{fig13}) is consistent with the presence
of a chiral condensate in the
absence of an infra-red fixed point.
One should keep in mind that one has to take the large $N$
limit at a fixed $b$ and then take the continuum limit, $b\to\infty$,
in order to arrive at a physically relevant conclusion.

\section{Future outlook}

We have provided compelling arguments for the preservation
of $Z_N^4$ symmetries in the large $N$ limit of Yang-Mills
theories coupled to adjoint overlap fermions on a single site lattice.
We have provided all the necessary details to perform a Hybrid
Monte Carlo algorithm and we have shown numerical feasibility.
We plan to study several different values of $N$ in the range of
$11$ to $18$ and we also plan to study several different couplings
in the range of $0.35$ to $5.0$~\cite{hnprep}. One can use a
physical observable like the Wilson loop to study the beta function
on the lattice and investigate the presence/absence of an infrared
fixed point. In this context,
it would
be useful to perform a weak coupling expansion of the beta
function on the single site lattice and confirm that one obtains the
same result as on an infinite lattice. Perturbative computations on
a single site lattice do not have the standard divergences and
a perturbative computation of the Wilson loop will
clarify 
the role of perimeter divergence on a single site lattice.
Finally, one should extend the numerical results to other
choices of boundary conditions that are not periodic.

\acknowledgments

We acknowledge discussions with Barak Bringoltz, Michael Buchoff, 
Aleksey Cherman, Aleksi Kurkela, Joyce Myers,
Herbert Neuberger, Erich Poppitz, and Mithat \"Unsal.
The authors acknowledge partial support by the NSF under grant number
PHY-0854744.  A.H also acknowledges partial support by the
U.S. DOE grant under Contract DE-FG02-01ER41172.

\end{document}